\documentstyle [12pt,epsfig,aaspp4] {article}

\newcommand{\be}{\begin{eqnarray}}
\newcommand{\ee}{\end{eqnarray}}
\def\lsim{\mathrel{\rlap{\lower4pt\hbox{\hskip1pt$\sim$}}
	\raise1pt\hbox{$<$}}} 
\def\gsim{\mathrel{\rlap{\lower4pt\hbox{\hskip1pt$\sim$}}
	\raise1pt\hbox{$>$}}} 

\def\kms{{km s$^{-1}$}}
\def\kpc{kpc}
\newcommand{\msun}{~{\rm M}_\odot}
\long\def\beginomit#1\endomit{}

\begin{document}

\title{THE FORMATION OF HIGH-MASS BLACK HOLES IN LOW-MASS X-RAY BINARIES}
\author{G. E. BROWN and C.-H. LEE}
\affil{Department of Physics and Astronomy,
        State University of New York\\
        Stony Brook, New York 11794-3800, USA}
\author{ HANS A. BETHE }
\affil{Floyd R. Newman Laboratory of Nuclear Studies,
        Cornell University \\
        Ithaca, New York 14853, USA}

\begin{abstract}
In this note we suggest that high-mass black holes; i.e., black holes of
several solar masses, can be formed in binaries with low-mass main-sequence
companions, provided that the hydrogen envelope of the massive star is
removed in common envelope evolution which begins only after the massive
star has finished He core burning. That is, the massive star is in the
supergiant stage, which lasts only $\sim 10^4$ years, so effects of mass
loss by He winds are small. Since the removal of the hydrogen envelope
of the massive star occurs so late, it evolves essentially as a single
star, rather than one in a binary. Thus, we can use evolutionary
calculations of Woosley \& Weaver (1995) of single stars.

Using the Brown \& Bethe (1994) upper limit of $\sim 1.8\msun$ for the
(gravitational) compact core mass that can evolve into a low-mass
black hole, we find that high-mass black holes can be formed in the
collapse of stars with ZAMS mass $\gsim 20\msun$. We somewhat arbitrarily
take the upper limit for the evolution of the so-called transient
sources to be $\sim 35\msun$ ZAMS mass. Mass loss by winds in stars
sufficiently massive to undergo the LBV (luminous blue variable) stage
may seriously affect the evolution of stars of ZAMS $>35-40\msun$, but
we need calculations with improved mass loss rates before discussing these
quantitatively. Both Portegies Zwart, Verbunt \& Ergma (1997) and
Ergma \& van den Heuvel (1998) have suggested that roughly our
chosen range of ZAMS masses must be responsible for the transient
sources. We believe that the high-mass black hole limit of ZAMS mass 
$\sim 40 \msun$ suggested by van den Heuvel \& Habets (1984) and
later revised to $\ge 50 \msun$ (Kaper et al. 1995) applies to
massive stars in binaries, which undergo RLOF (Roche Lobe Overflow) early
in their evolution. We will not pursue this here because calculations with
improved He-star mass loss rates by wind, now being carried out,
are necessary before
quantitative results can be obtained.

The most copious high-mass black holes of masses $\sim 6 -7 \msun$
have been found in the transient sources such as A0620.
These have low-mass companions, predominantly of $\lsim 1 \msun$, such as
K-- or M--stars.
In the progenitor binaries the mass ratios must have
been tiny, say $q\sim 1/25$. Normally such small ratios are thought
to be rare; e.g. in binary evolution the companion distribution
is often taken as $dq$, implying an very low probability of
such a binary.

In this note we follow the evolutionary scenario for the black hole
binary of de Kool et al. (1987). We show
that the reason for this small $q$-value
lies in the common envelope evolution of the binary. The smaller
the companion mass, the greater the radius $R_g$ the giant must reach
before its envelope meets the companion.
This results because 
the orbit of a low-mass companion must shrink by a large factor in
order to expel the envelope of the giant, hence the orbit must initially
have a large radius. (Its final radius must be just inside its Roche
Lobe, which sets a limit to the gravitational energy it can furnish.)

A large radius $R_g$ in turn means that the primary star must be in the
supergiant stage. Thus it will have completed its He core burning
while it is still ``clothed" with hydrogen. This prevents excessive
mass loss so that the primary retains essentially the full mass of its
He core when it goes supernova.
We believe this is why
K-- and M--star companions of high-mass black holes are favored.

We find that the black holes in transient sources can be formed
from stars with ZAMS masses in the interval $20-35\msun$.
The black hole mass is only slightly smaller than the He core mass,
typically $\sim 7\msun$.

\end{abstract}

\setcounter{equation}{0}
\section{DEVELOPMENT}
\label{sec1}

Our evolutionary scenario is essentially the same as that of
de Kool et al. (1987) for the black hole binary A0620-00. 
We apply this scenario somewhat schematically for the range of
ZAMS $20-35\msun$ for the massive star. In Fig.~\ref{fig1}
we show results of calculations of Fe core masses of single
stars by Woosley \& Weaver (1995). Somewhere around ZAMS mass
$20\msun$ the Fe core exceeds the Brown \& Bethe (1994) limit
of $\sim 1.8\msun$ for low-mass black-hole formation, so we take this
as the beginning of high-mass black hole formation.
The detailed behavior of this curve should not be taken seriously,
but the large increase around $20-25\msun$ is of importance, as
we discuss in detail later.

We wish to carry out a population synthesis so that we can estimate
the number of transient sources.
We make roughly the same assumptions as Bethe \& Brown (1998,1999),
our massive star $M_B$ lying in mass somewhere in between the
O,B star progenitors of binary neutron stars and the progenitor
of the massive black hole in Cyg X-1. Our low-mass companion is
a main sequence star of mass $\sim 1 \msun$. Thus, the ratio
\be
q= \frac{M_{Ai}}{M_{Bi}}
\label{2.1}
\ee
is very small, and there will be great uncertainty in the initial
number of binaries for such a small $q\sim 1/25$. We take
the distribution as $dq$.
The distribution in $q$ is unknown for such low-mass companions
as are involved here, but our results will show that the flat
distribution in $q$ is not unreasonable.
We assume $\log a$, where $a$ is the
semi-major axis of the orbit, to be uniformly distributed, over
a total logarithmic interval of 7. Thus, the fraction of binaries in a
given interval of $\ln a$ is
\be
d\phi = \frac{d(\ln a)}{7}.
\label{2.2}
\ee
We take $\alpha$ the supernova rate to be $\sim 0.02$ per galaxy per year,
somewhat larger than the 0.015 given by Cappellaro et al. (1997) and 
somewhat smaller
than the 0.025 assumed by Bethe \& Brown (1998). The rate of supernovae (SN)
is the same as the rate of birth of massive stars $M> 10 \msun$, so
\be
d\alpha =\alpha n\left(\frac{M}{10 \msun}\right)^{-n}\frac{dM}{M}
\ee
with $n$ the Salpeter exponent which we take to be $1.5$. We furthermore
assume a binarity of 0.5. Once again, for our very small $q$ values
this is uncertain.

We thus evolve as typical 
a ZAMS $25 \msun$ (star B) with companion $\sim 1 \msun$
main sequence star (star A) as typical progenitor of the transient X-ray
sources. The common envelope evolution can be done as in Bethe \& Brown (1998).
With $M_{B_i}=25 \msun$
and neglect of the accretion onto
the main sequence mass $M_A$, we find from Bethe \& Brown
\be
\left(\frac{Y_f}{Y_i}\right)^{1.2} =
   \frac{1.2}{\alpha_{ce}}  \frac{ M_{B_i}}{M_A}
\ee
where $Y=M_B/a$. Here the coefficient of dynamical friction $c_d$ was
taken to be 6, in the range of the 6-8 for supersonic flow with Mach
number 3-10 (Ruffert 1994; Ruffert \& Arnett 1994).
The result is relatively insensitive to $c_d$, the exponent
$1.2$ resulting from $1+1/(c_d-1)$.

Thus, in our case
\be
\frac{Y_f}{Y_i} =
   17 \left(\frac{\alpha_{ce} M_A}{M_\odot}\right)^{-0.83}
  =
  30 \left(\frac{0.5}{\alpha_{ce}}\frac{M_\odot}{M_A}\right)^{0.83}.
\ee
We expect $\alpha_{ce} \simeq 0.5$, under the assumption
that the kinetic energy of the expelled envelope is equal to that
it originally possessed in the massive star, but it could be smaller.
{} From this we obtain
\be
\frac{a_i}{a_f} = \frac{M_{B_i}Y_f}{M_{B_f}Y_i} =
   90 \left(\frac{0.5}{\alpha_{ce}}\frac{M_\odot}{M_A}\right)^{0.83},
\ee
where we have taken the He star mass $M_{B_f}$ to be $1/3$ of
$M_{B_i}$.
In order to survive spiral-in, the final separation $a_f$ must
be sufficient so that the main sequence star lies at or inside its
Roche Lobe, about $ 0.2 a_f$ if $M_A = \msun$.
This sets $a_f\sim 5 R_\odot = 3.5\times 10^{11}$~cm and
\be
a_i = 3.15
   \left(\frac{0.5}{\alpha_{ce}}\right)^{0.83}
  \times 10^{13}\; {\rm cm}.
\label{4.7}
\ee
$\sim 2$ A.U..
This exceeds the radius of the red giant tip in 
the more numerous lower
mass stars in our interval, so the massive star must generally be in the {\em
supergiant} phase where it meets the main sequence star, i.e., the massive star
must be beyond He core burning.  E.g., the red giant tip (before the He core
burning)
for a $20\msun$ star is at
$0.96\times 10^{13}$~cm, for a $25\msun$ star, $2.5\times 10^{13}$~cm
(Schaller et al. 1992).
These numbers are, however, somewhat uncertain.
Notice that decreasing $\alpha_{ce}$ will
increase $a_i$.
Decreasing $M_A$ has little influence, because
with the smaller stellar radius the minimum $a_f$ will decrease nearly
proportionately.
Note that neglect of accretion onto the main sequence
star would change the exponent $0.83$ to unity, so accretion is
unimportant except in increasing the final mass.

Now a ZAMS $25 \msun$ star ends up at radius $6.7\times 10^{13}$~cm
($\sim 2~a_i$)
following He shell burning (Weaver, Zimmerman \&Woosley 1978).
Thus the interval between $a_i$ and $6.7\times 10^{13}$~cm
is available for spiral-in without merger\footnote{
Note that envelope removal does not occur at the Roche Lobe on the
thermal time scale $\tau_{th}$ but at the low-mass star since
the remaining lifetime of the giant
is $\sim 10^4$ yrs, much shorter than $\tau_{th}$.
}
so that a fraction
\be
\frac{1}{7} \ln\left( \frac{6.7}{3.15
   \left(\frac{0.5}{\alpha_{ce}}\right)^{0.83}}\right)
\simeq 0.11
\label{4.8}
\ee
of the binaries survive spiral-in, but are close enough so that the
main sequence star is encountered by the evolving H envelope of the
massive star.
The He core burning will be completed before the supergiant has moved
out to $\sim 2$ A.U., so binaries which survive spiral-in
will have He cores which burn as ``clothed", namely as in single stars.

\setcounter{equation}{0}
\section{BIRTH RATE OF TRANSIENT SOURCES}
\label{sec2}

Given our assumptions in Section \ref{sec1}, 
the fraction of supernovas which arise
from ZAMS stars between 20 and $35 \msun$ is
\be
\frac{1}{2^{3/2}}-\frac{1}{3.5^{3/2}} =0.20
\label{3.1}
\ee
where we have assumed the mass $10 \msun$ is necessary for a star
to go supernova. 
A Salpeter function with index $n=1.5$ is assumed here.
Our assumption that the binary distribution is as $dq$
is arbitrary, and gives us a factor $1/25$ for a $1 \msun$
companion. Note, however, that had we included higher mass companions,
the change in the final projected number of transient sources would not be
order of magnitude, because the hydrogen burning time goes inversely
as mass squared. Thus, for supernova rate $2$ per century, our birth rate
for transient sources in the Galaxy is
\be
2\times 10^{-2}\times 0.5\times 0.11 \times 0.20 \times 0.04
\simeq 8.8\times 10^{-6} {\rm yr}^{-1}
\label{3.2}
\ee
where $0.5$ is the assumed binarity, $0.11$ comes from eq.~(\ref{4.8}),
and 
the final (most uncertain) factor $0.04$ results from a
distribution flat in $q$ and an assumed $1\msun$ companion star.

In order to estimate the number of transient sources with black holes
in the Galaxy, we should know the time that a
main sequence star of mass $ \sim 1\msun$ transfers mass to a more
massive companion. For a main-sequence donor, the mass transfer rate is
$\sim 10^{-10}\msun {\rm yr}^{-1}$, almost independent of donor mass
(Verbunt \& van den Heuvel 1995). As mass is transferred, the mass of
the donor decreases and with it the radius of the donor.
Quite a few low-mass X-ray binaries have X-ray luminosities that imply
accretion rates in excess of $10^{-10}\msun {\rm yr}^{-1}$, leading to
suggestions of additional mechanisms for loss of angular momentum from
the binary, to increase mass transfer. Verbunt \& Zwaan (1981) estimate
that magnetic braking can boost the transfer of mass in a low-mass binary.
We somewhat arbitrarily adopt an effective mass transfer rate of
$10^{-9} {\rm yr}^{-1}$ for main sequence stars and $10^{-8} {\rm yr}^{-1}$
for the two systems that have subgiant donors (V404 Cyg and XN Sco 94).
In order to estimate the number of high-mass black hole, main sequence
star binaries in the Galaxy we should multiply the birth rate 
(\ref{3.2}) times the $10^9$ yrs required, at the assumed mass loss rate,
to strip the main sequence star, obtaining
8800 as our estimate. Not all of these will fill their Roche Lobes.
Those that do not may not now be visible, but will be later, as they 
begin evolving as one of those with subgiant donors.
The fact that two of seven observed binaries are subgiants, although
the lifetime of the latter is two orders of magnitude less than the main
sequence lifetime, suggests that some fraction of our 8800 estimated
binaries do not fill their Roche Lobes.
From the observed black-hole transient sources Wijers (1996) arrives
at 3000 low-mass black hole sources in the Galaxy, but regards
this number as a lower limit. 
Beginning from this, which we regard as an observational estimate,
we note that the two subgiants in Table 1 involve the more massive
F,G and A stars. These indicate an $\sim 100$ times greater population
of unevolved main sequence stars in this range which lie quietly inside
their Roche Lobes. Thus, including the quiescent binaries might give
as many as $(2/7)\times 100\times 3000$ or $\sim 10^5$ additional
binaries, suggesting that our above estimate may be an order of magnitude too
low.
Estimates of the number of transient
sources are very uncertain, but it is clear that there are orders of
magnitude more of them than of the Cygnus X-1 type objects with
high-mass black hole and massive star companion. 

If we assume that ZAMS masses $\sim 10 -18 \msun$ evolve into a neutron star,
we should have $\sim 3$ times more neutron stars than high-mass black holes
(see eq.~(\ref{3.1})). The upper limit follows from our belief that SN (1987A)
with progenitor $\sim 18\msun$ ZAMS went into a low-mass black hole, following
the scenario of Brown \& Bethe (1994). On the basis of a Monte Carlo
calculation using the kick velocities of Cordes \& Chernoff (1997) we
find that $\sim 1/2$ of the binaries containing He-star, low-mass main sequence
companion (with $M\simeq 1 \msun$) will be disrupted in the explosion. Thus, we
find only a slightly higher birth rate for LMXB's with neutron stars than with
black holes, although the numbers could be equal to within our accuracy. The
LMXB's with neutron stars tend to be much brighter than those with black holes,
indicating an order of magnitude greater transfer rate. With the
correspondingly shorter main-sequence lifetime, this would give us several
hundred LMXBs with neutron stars, a factor of several greater than the observed
number, $\sim 130$. Given the lifetime of $\sim 10^8$ years of a LMXB
and accretion at roughly the Eddington rate of 
$\sim 10^{-9}-10^{-8}\msun$ yr$^{-1}$,
it is reasonable that some neutron stars accrete a reasonable fraction
of a solar mass $\msun$ (van den Heuvel 1995). We expect the masses of
these to exceed the Brown \& Bethe (1994) limit of $1.5\msun$ for
maximum neutron star mass,
and evolve into low-mass black holes. The low-mass black-hole, low-mass
main sequence star systems would not be seen.

\setcounter{equation}{0}
\section{ESTIMATED MASSES OF THE BLACK HOLES IN TRANSIENT SOURCES}
\label{sec3}

As we showed below eq.~(\ref{4.7}), 
the He core of the massive star will in general
be uncovered only after He core burning is completed.  The remaining time for
He burning (in a shell) will be short, e.g., for a $20 \msun$ ZAMS star it is
only $1.4 \times 10^4$ years (Schaller et al. 1992).  Therefore the mass loss by
wind after uncovering the He core will not be large, and when the star finally
becomes a supernova, its mass will be almost equal to the He core of the
original star.  The latter can be calculated from
\be
M_{He} \simeq 0.08 (M_{ZAMS})^{1.4}
\label{eq:6.1}
\ee
so for ZAMS masses $20-35\msun$ $M_{He}$ will lie in the interval
$\sim 5.3 -11.6\msun$. In fact, the lower limit looks a bit small,
because the He core of the $\sim 18\msun$ progenitor of 1987A is generally
taken as $\sim 6\msun$.

Bailyn et al. (1998) find the black hole masses in transient sources
to be clustered about $\sim 7\msun$, except for V404 Cyg which has a
higher mass. This is in general agreement with our scenario, because
most of the black holes will come from the more numerous stars of ZAMS
mass not far from our lower limit of $\sim 20\msun$. 
Two points are important to note:

\begin{enumerate}
\item Not much mass can have been lost by wind. Naked He stars have
rapid wind loss. 
However in our scenario the He star is made naked only
during He shell burning and therefore does not have much time
($\lsim 10^4$ yrs) to lose mass by wind.
\item There are good reasons to believe that the initial He core will be
rotating (cf. Mineshige, Nomoto \& Shigeyama 1993). The way in which
the initial angular momentum affects the accretion process has been studied
by Mineshige et al. (1997) for black hole accretion in supernovae.
In general accretion discs which are optically thick and advection
dominated are formed. The disc is hot and the produced energy and photons
are advected inward rather than being radiated away. The disc material
accretes into the black hole at a rate of $> 10^6\dot M_{Edd}$ for
the first several tens of days. Angular momentum is advected outwards.
Our results show that little mass is lost, because the final
$\sim 7\msun$ black hole masses are not much less massive than the He core
masses of the progenitors, and some mass is lost by wind before the
core collapses.
The latter  loss will not, however, be great, because there is not much
time from the removal of the He envelope until the collapse.
\end{enumerate}

Accretion of the He into the black hole will differ quantitatively
from the above, but we believe it will be qualitatively similar.
The fact that the helium must be advected inwards and that little mass is lost
as the angular momentum is advected outwards is extremely
important to establish. This is because angular momentum, essentially
centrifugal force, has been suggested by Chevalier (1996) to hold
up hypercritical accretion onto neutron stars in common envelope evolution.
(Chevalier (1993) had first proposed the hypercritical accretion during
this evolutionary phase to turn the neutron stars into black holes,
the work followed up by Brown (1995) and Bethe \& Brown (1998).)
However, once matter is advected onto a neutron star, temperatures
$\gsim 1$ MeV are reached so that neutrinos can carry off the energy.
The accreted matter simply adds to the neutron star mass, evolving
into an equilibrium configuration. Thus, this accretion does not differ
essentially from that into a black hole.
In either case of neutron star or black hole an accretion disc or
accretion shock, depending on amount of angular momentum, but both
of radius $\sim 10^{11}$~cm, is first formed, giving essentially
the same boundary condition for the hypercritical accretion in
either case, black hole or neutron star. Thus, the masses of the
black holes in transient sources argue against substantial inhibition
of hypercritical accretion by jets, one of the Chevalier (1996) suggestions.

Measured mass functions, which give a lower limit on the black hole mass are
given in Table~\ref{tab1}. Only GRO J0422+32 and 4U 1543-47 have a measured
mass function $\lsim 3  \msun$.
Results of Callanan et al. (1996) indicate that the angle $i$ between
the orbital plane and the plane of the sky for GRO J0422+32 is $i<45^\circ$,
and recent analysis by Orosz et al. (1998) indicate that the angle $i$ for 4U 1543-47
is $20^\circ <i< 40^\circ$.
So both GRO J0422+32 and 4U 1543-47 also contain  high-mass black holes.

\setcounter{equation}{0}
\section{GENERAL DISCUSSION}
\label{sec4}

There is agreement (Portegies Zwart et al. 1997; Ergma \& van den Heuvel, 1998)
that in order to make enough transient sources the progenitors of the black 
holes must begin at relatively low masses, $\sim$ ZAMS $20\msun$.
We take the upper limit somewhat arbitrarily to be $\sim 35\msun$ about
where rapid mass loss occurs and stars may enter into the LBV phase.
Our upper limit is relatively unimportant since most of the stars considered
will lie near the lower limit.

Based on the observations of Kaper et al. (1995) that the companion
is a hypergiant, Ergma \& van den Heuvel (1998) argue that the progenitor of
the neutron star in 4U1223-62 must have a ZAMS mass $\gsim 50\msun$.
Brown, Weingartner \& Wijers (1996), by similar argumentation, arrived
at $\sim 45\msun$, but then had the difficulty that 4U1700-37,
which they suggested to contain a low-mass black hole appeared to evolve
from a lower mass star than the neutron star in 1223. Wellstein \& Langer (1999)
suggest the alternative that in 1223 the mass occurs in the main sequence
phase (Case A mass transfer) which would be expected to be quasi conservative.
They find that the progenitor of the neutron star in 1223 could come from
a mass as low as $26\msun$. This is in agreement with Brown et al. (1996)
for conservative mass transfer (their Table 1), but these authors
discarded this possibility, considering only RLOF (Case B mass transfer)
in which case considerable mass would be lost.

Wellstein \& Langer (1999) are in agreement with Brown et al. (1996) that
4U1700-37 should come from a quite massive progenitor.
Conservative evolution here is not possible because of the short period
of 3.4 days (Rubin et al. 1996). The compact object mass is here
$1.8 \pm 0.4\msun$ (Heap \& Corcoran 1992). Brown et al. (1996) suggest
that the compact object is a low-mass black hole. The upper mass limit
for these was found by Brown \& Bethe (1994) to be $\sim 1.8\msun$,
as compared with an upper limit for neutron star masses of $\sim 1.5\msun$.
Thus, there seems to be evidence for some ZAMS masses of $\sim 40-50\msun$
ending up as low-mass compact objects, whereas we found that lower mass
stars in the interval from $\sim 20-35\msun$ ended up as
high-mass black holes. In this sense we agree with Ergma \& van den Heuvel
(1998) that low-mass compact object formation ``is connected with
other stellar parameters than the initial stellar mass alone." We suggest,
however, following Brown et al. (1996) that stars in binaries evolve
differently from single stars because of the different evolution of the
He core in binaries resulting from RLOF in their evolution. Namely,
``naked" He cores evolve to smaller final compact objects than ``clothed"
ones. 

\begin{figure}[ht]
\centerline{\epsfig{file=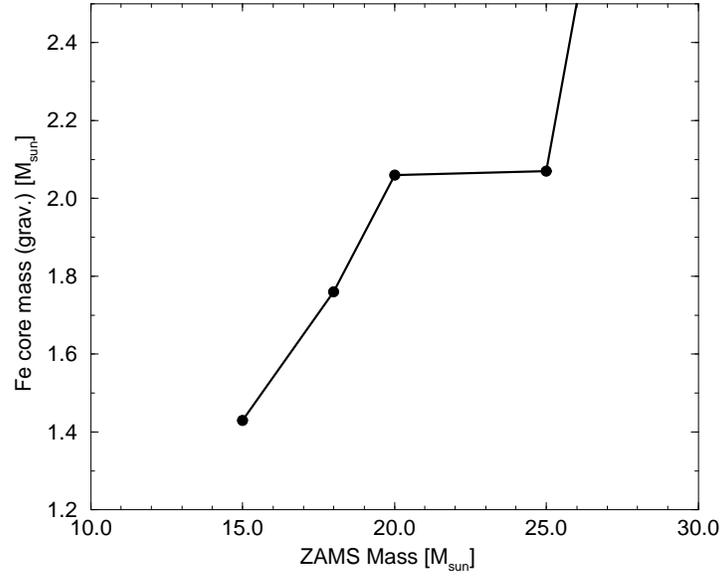,height=9cm}}
\caption{Comparison of the compact core (Remnant) masses resulting from the
evolution of single stars (filled symbols), case of solar
metallicity of Woosley \& Weaver (1995).
}
\label{fig1}
\end{figure}

In fact, this different evolution of binaries was found by Timmes, Woosley
\& Weaver (1996). They pointed out that stars denuded of their hydrogen
envelope in early RLOF in binaries would explode as Type Ib supernovae.
They found the resulting remnant gravitational mass following explosion
to be in the interval of $1.2-1.4\msun$, whereas in exploding stars of all
masses with hydrogen envelope (Type II supernova explosion) they found
a peak at about $1.28\msun$, chiefly from stars of low masses and another
peak at $1.73\msun$ more from massive stars. Our Fe core masses in
Fig.~\ref{fig1} come from essentially the same calculations, but the
``Remnant" masses
of Woosley \& Weaver (1995) are somewhat greater than those used
by Timmes et al. (1996). In fact, the differences between the masses we
plot and those of Timmes et al. come in the region $\sim 1.7-1.8\msun$
(gravitational). This is just in the Brown \& Bethe (1994) range for 
low-mass black holes. It may be that some of the stars with low-mass
companions evolve into low-mass black holes. Presumably
these would give lower luminosities than the high-mass black holes, although
at upper end of the mass range we discuss 4U1700-37 seems to be an
example of such a system.
Of course here the high luminosity results from the high mass loss rate
of the giant companion.
There are substantial ambiguities in fallback, etc., from the explosion.
Our point in this paper is that most of the higher mass single stars 
$20-35\msun$ go into high mass black holes. (The Brown \& Bethe (1994)
limit for low-mass black hole formation is $\sim 1.5-1.8\msun$ gravitational,
but there is some give and take in both lower and upper limit.
Also the stars are not all the same. In particular different metallicities
will give different wind losses.)

In determining the upper mass limit for which a low-mass compact object can
result from binary evolution with RLOF (Case B mass transfer) the Brown et al.
(1996) scenario must be revised because it is now realized that He-star wind
loss rates employed by Woosley, Langer \& Weaver (1995) were too high.
With lower rates the differences in behavior between ``clothed"" and ``naked"
He core evolutions will be diminished because much of the difference arises
from the mass losses and how it affects the convective $^{12}C$ burning
as we discuss below.

In the mass range $\sim 20-35\msun$, the compact objects resulting from naked
He stars (Type Ib SN explosions) are sufficiently far below the maximum
neutron star mass that they will still remain neutron stars when better (lower)
He star mass loss rates are employed. Indeed, for our $25\msun$ star, 
Wellstein and Langer (1999) find that halving the mass loss rate lowers the
mass of the CO core by only 2.5 \%.
However, at about the upper end of this range of masses, the outcome may,
indeed, be changed. We will somewhat arbitrarily focus on the mass region
$\sim 40\msun$, studying how lower mass loss rates might affect the outcome.

Until recently W.-R. wind loss rates were taken from observed winds which
originated chiefly from free-free scattering. These depend 
quadratically on density. Because of ``clumpiness" in the winds, the
mass loss rate was overestimated. Polarization measurements of the
Thomson scattering, which depend linearly on the wind, give substantially
lower mass rates, in approximate agreement with the rates that would
be deduced from the observed rate of increase in orbital periods for
spherical mass loss
\be
\frac{\dot M}{M}=
\frac{2 \dot P}{P} .
\label{neq.4.1}
\ee
In V444 Cygni $\dot P=0.202\pm 0.018\; s \; {\rm yr}^{-1}$
was obtained by Khaliullin et al. (1984) and
$M_{WR} =9.3\pm 0.5 \msun$ by Marchenko et al. (1994), resulting in 
\be
\dot M_{dyn} = 1.03\times 10^{-5} \msun {\rm yr}^{-1}.
\label{neq.4.2}
\ee
This is to be compared with the
\be
\dot M=0.75\times 10^{-5}\msun {\rm yr}^{-1}
\ee
obtained by St.-Louis et al. (1993).
In later work Moffat \& Robert (1994) arrive at a mean of 
$(0.7\pm 0.1)\times 10^{-5}\msun {\rm yr}^{-1}$.
The mass loss rate employed by Woosley, Langer \& Weaver (1995) was
that of Langer (1989). Specifically, $\dot M=-k M^{2.5}$,
with $M$ in $\msun$ and $\dot M$ in $\msun {\rm yr}^{-1}$ and
$k=6\times 10^{-8}$ as long as the carbon surface mass fraction does not
exceed $0.02$ and $k=10^{-7}$ afterwards. Choosing an average of
$k=8\times 10^{-8}$ we find the WLW rate to be $2.1\times 10^{-5}\msun 
{\rm yr}^{-1}$, a factor of 2 larger than $\dot M_{dyn}$. Given the
many uncertainties in our estimate, we feel the range of two to three times
less than the Langer (1989) mass loss rate to be reasonable.
It should also be remembered that stars vary substantially in metallicity,
and that there will be a range of variation even of those in the Galactic
disc. Of course stars in the metal poor Magellanic clouds should have
substantially lower winds.

In the region of ZAMS $\sim 40\msun$ calculation of Wellstein \& Langer
(1999) show that both the final He core mass and C/O core mass increase
$\sim 23\%$ in the mass loss rate is halved. Thus, reducing the mass loss
rate by a factor of 2 does not increase the final mass by a similar
factor. The main reason is that if the mass loss is reduced, then the
He stars remain somewhat more massive, thus also more luminous and therefore
have a higher mass loss rate than had their mass been reduced earlier.
There are also other, less important, feedbacks.

In the case of $\sim 60\msun$ stars, the He core and C/O cores increase
$\sim 31\%$ if the mass loss rate is halved and a factor of $\sim 1.8-1.9$
if it is cut by $1/4$. In fact, in this case, a reasonable interpolation
formula for these lower mass loss rates is
\be
\frac{M}{M_0}=(1.33)^{1/2f}
\ee
where $M_0$ was calculated with the Langer (1989) rates and
$f$ is the fractional decrease in winds from these rates.
Wellstein \& Langer (1999) also evolve a $36\msun$ He core, which
would come from an $\sim 85\msun$ ZAMS star, with $f=1/4$ of the
Langer et al (1989) wind loss rates.
Their final $7.5\msun$ He star mass is not enough for Cyg X-1.
However, Woosley, Langer \& Weaver (1993) with the Langer
1989 mass loss rate found a $9.71\msun$ He core in evolving
a ZAMS $85\msun$ star with mass loss. In the latter case the
WNL phase lasted $>1/3$ of the W.-R. phase; i.e., the He core
initially had some hydrogen envelope. This illustrates the order
of uncertainties that may come in the evolution of very massive
stars, arising from mass loss in the LBV phase.

Although the final He core and C/O core scale by roughly the same factor
the Fe core, crucial for the compact object masses, is not expected
 to do the same.
Firstly, rather trivially, even if the baryon number Fe core did scale
in the same way, one would expect corrections downwards in core mass
from binding energy correction. These decreases in gravitational mass
would be greater for higher mass cores, so the Fe cores would scale
somewhat less rapidly than the C/O cores.

More important for the mass of the Fe core is the ZAMS mass at which
the convective carbon burning is skipped, because, as seen in the single
stars in Fig.~\ref{fig1}, a big jump in Fe core masses occurs here.

The convective carbon burning phase (when it occurs) is
extremely important in presupernova evolution,
because this is the first phase in which a large amount of entropy
can be carried off in $\nu\bar\nu$-pair emission, especially because
this phase is of long duration.  The reaction in which carbon burns is
$^{12}C(\alpha,\gamma)^{16}O$ (other reactions like $^{12}C+$$^{12}C$ 
would require
excessive temperatures).  The cross section of $^{12}C(\alpha,\gamma)^{16}O$ is
still not accurately determined; the lower this cross section the higher the
temperature of the $^{12}C$ burning, and therefore the more intense the
$\nu\bar\nu$ emission.
With the relatively low
$^{12}C(\alpha,\gamma)^{16}O$ rates\footnote{ 
Weaver \& Woosley use $S(E)=S(300)=170$ keV barns, remarkably close
to the $169\pm 55$ keV barns arrived at by Barnes (1995).
Given the large uncertainty (stemming chiefly from that in the E2 rate,
the good agreement may be somewhat accidental.
}
 determined both directly from nuclear
reactions and from nucleosynthesis by Weaver \& Woosley (1993), the
entropy carried off during $^{12}C$ burning in the stars of ZAMS
mass $\le 18 \msun$ is substantial.
The result is rather low-mass Fe cores for these stars, which can evolve
into neutron stars. Note that in the literature earlier than
Weaver \& Woosley (1993) often large $^{12}C(\alpha,\gamma)^{16}O$
rates were used, so that the $^{12}C$ was converted into oxygen and the
convective burning did not have time to be effective.
Thus its role was not widely appreciated.

Of particular importance is the ZAMS mass at which the convective carbon
burning is skipped.
In the Woosley \& Weaver (1995) calculations this occurs for single stars at
ZAMS mass $19\msun$ but with a slightly lower $^{12}C(\alpha,\gamma)^{16}O$
rate it might come at $20\msun$ or higher (Brown 1997). As the progenitor
mass increases, it follows from general polytropic arguments that the
entropy at a given burning stage increases.
At the higher entropies of the more massive stars the density at which
burning occurs is lower, because the temperature is almost fixed for a
given fuel. Lower densities decrease the rate of the triple-$\alpha$
process which produces $^{12}C$ relative to the two-body
$^{12}C(\alpha,\gamma)^{16}O$ which produces oxygen.
Therefore, at the higher entropies in the more massive stars the
ratio of $^{12}C$ to $^{16}O$ at the end of He burning is lower.
The star skips the long convective carbon
burning and goes on to the much shorter oxygen burning.  
Oxygen burning goes via $^{16}O +$$^{16}O$ 
giving various products, at very much higher temperature
than $^{12}C(\alpha,\gamma)^{16}O$ and much faster.
Since neutrino cooling during the long carbon-burning phase gets
rid of a lot of entropy of the core, skipping the convective carbon 
burning phase leaves
the core entropy higher and the final Chandrasekhar core fatter.

In Fig.~\ref{fig1} the large jump in compact object mass in single stars at
ZAMS mass $\sim 19\msun$ is clearly seen. From our discussion in
the last section we see that this is just about at the point where
our Fe core  mass goes above $\sim 1.8\msun$ and,
therefore, above this mass one would expect single stars to go into
high-mass black holes. Arguments have been given that SN (1987A)
with progenitor ZAMS mass of $\sim 18 \msun$ evolved into a low-mass
black hole (Brown \& Bethe 1994). We believe from our above arguments
and Fig.~\ref{fig1} that soon above the ZAMS mass of $\sim 18\msun$, single
stars go into high-mass black holes without return of matter to the
Galaxy.
Thus, the region of masses for low-mass black hole formation in
single stars is narrow.
The precise  upper mass limit is not
clear, but certainly in the range of $\gsim 20\msun$ ZAMS.
This is in agreement with Wellstein \& Langer (1999) who find a minimum
black hole progenitor mass for single stars of $21\msun$.

Thus far our discussion has been chiefly about single stars, in which the 
He burns while clothed by a hydrogen envelope.

In binary evolution, if the hydrogen envelope is removed before the helium
core burning in either Case A or Case B (RLOF) mass transfer, the resulting
``naked" He star burns quite differently from a ``clothed" one.
Weaver \& Woosley (1993) find that the convective carbon burning tends to
be skipped when the central $^{12}C$ abundance at the end of helium
core burning is less than $\sim 15\%$. In Woosley, Langer \& Weaver (1995)
with the Langer (1989) helium mass loss rate, this central $^{12}C$ abundance
is $34\%$ for a ZAMS $40\msun$ star. With mass loss rate decreased to half,
it is still $33\%$ (Wellstein \& Langer, 1999) hardly changed. 
However, as noted earlier, the CO mass for a ZAMS $40\msun$ star is
increased from $2.33\msun$ to $2.87\msun$ with halved mass loss rate.
Provisionally Wellstein \& Langer (1999) have suggested that the
magnitude of the CO mass chiefly determines the fate of the star, and
that the $2.87\msun$ CO core could go into a black hole. However, halving
the mass loss rate of the naked He star decreases the central carbon abundance
at the end of He core burning hardly at all.
Thus, there will still be a long period of $^{12}C$ convective core 
burning. As seen from Fig.~\ref{fig1} for less massive stars in the mass
region $\sim 20\msun$, this can easily lower the compact core mass by 
$\sim 0.5 \msun$. On the other hand, for each $1\msun$ added to the
He envelope at the envelope at the time of SN explosion, an  additional
 $\sim 3\times 10^{49}$ ergs is needed to expel it in either a prompt or
delayed supernova explosion. This is $\sim 2\%$ of the SN explosion energy,
so would work towards formation of a high-mass black hole.

Given the increase in He and CO cores from the decreased mass loss rates,
we estimate that stars of ZAMS masses $\sim 40\msun$ which lose their
H envelopes by wind will end up as low-mass black holes, 1700-37 being
an example. The possible mass ranges in which this can happen cannot be
determined until the dynamic evolution of the CO cores formed
with the lowered He-star wind rates of Wellstein \& Langer (1999)
is carried out.

It seems clear that with the lower metallicity in the LMC and consequently lower
mass loss rates that the $^{12}C$ convective core burning will not be skipped
so the Fe cores in the $\gsim 40\msun$ region of ZAMS masses will
be larger, and will evolve into black holes. This may help to explain why
there are two HMXB's with high-mass black holes, LMC X-1 and LMC X-3,
in the LMC, whereas Cyg X-1 is the only clear example in the disc. 

\setcounter{equation}{0}
\section{CONCLUSION}

We have shown that it is likely that single stars in the range of ZAMS
masses $\sim 20-35\msun$ evolve into high-mass black holes without return of
matter to the Galaxy. This results because at mass $\sim 20\msun$ the
convective carbon burning is skipped and this leads to substantially
more massive Fe cores. Even with more realistic reduced mass loss
rates on He stars, it is unlikely that stars in this mass range in binaries
evolve into high-mass black holes,
because the progenitor of the compact object when stripped of its
hydrogen envelope in either Case A (during main sequence)
 or Case B (RLOF) mass transfer will burn as a
``naked" He star, ending up as an Fe core which is not sufficiently
massive to form a high-mass black hole.

In the region of ZAMS mass $\sim 40 \msun$, depending sensitively on
the rate of He-star wind loss, the fate of the primary in a binary may
be a low-mass black hole.
We are unable to pin down the limit for high-mass black hole formation
until better mass loss rates are determined.

In our estimates we have assumed the Brown \& Bethe (1994) estimates of
$1.5\msun$ for maximum neutron star mass and $1.5-1.8\msun$ for the range
in which low-mass black holes can result.

In our evolution of the transient sources using Case C (during He shell 
burning) mass transfer, almost
the entire He core will collapse into a high-mass black hole, explaining
the more or less common black hole mass of $\sim 7\msun$ for these objects,
with the possible exception of V404 Cygni where the mass may be greater.
Our evolution gives an explanation for the seemingly large gap in
black-hole masses, between the $\gsim 1.5\msun$ for the black hole we believe
was formed in 1987A and the $\sim 1.8\msun$ black hole we suggest in
1700-37 and the $\sim 7\msun$ in the transient sources.

We note that following the removal of the H envelope by Case C mass
transfer, the collapse inwards of the He envelope into the developing
black hole offers the Collapsar scenario for the most energetic
gamma ray bursters of Woosley (1993) and MacFadyen \& Woosley (1999).
Especially if our higher estimate of $\sim 10^5$ of the high-mass black-hole,
main-sequence star binaries is roughly correct, a simple estimate
(Brown, Lee \& Wijers 1999) shows that they would be the largest population
of possible progenitors. 

\acknowledgements

We dedicate this work to Dave Schramm, a giant among astrophysicists.
One of the authors (G.E.B.) invited him to edit the astrophysics
section of Physics Reports and enjoyed his collegiality for many years.

We are especially grateful to S. Wellstein and N. Langer for communicating
their results with modified mass loss rates to us before publication.
We would like to thank Charles Bailyn for useful discussions and, especially,
for putting the data in Table 1 together for us.
We would like to thank Stan Woosley, who not only provided us with
most of the results we used, but also clarified their interpretation
in several communications. We also thank Adam Burrows for a clear
explanation of the consequences of skipping the convective carbon
burning.
We wish to thank Simon Portegies Zwart for several helpful discussions.
We were supported by the U.S. Department of Energy under Grant No.
DE--FG02--88ER40388.

\newpage

\def\arraystretch{0.7}

\begin{table}[ht]
   \caption[Black holes candidates with measured mass functions]{Parameters
	   of suspected black hole binaries with measured mass functions
           (Wijers 1996, Chen et al. 1997, Bailyn et al. 1998, Orosz et al. 1998, 
            Bailyn, private communication).
	   N means nova, XN means X-ray nova. Numbers in parenthesis indicate
	   errors in the last digits.
           }
   \label{tab1}
\begin{center}
\small
\newcommand{\ti}[1]{{\tiny #1}}
\noindent\begin{tabular}{@{}llccccc@{}}\hline
                &               &  compan. &$P_{orb}$   &$f(M_{X})$&$M_{opt}$ & $(l,b)$ \\
X-ray           & other         &  type    & (d)        &  ($\msun$)    &  ($\msun$)    & \\ \cline{3-7}
names           & name(s)       &  q       &$K_{opt}$   &  i          &$M_{X}$   &    $d$ \\
                &               &  ($M_{opt}/M_X$) & (\kms)     &  (degree)   &  ($\msun$)    &  (\kpc) \\ \hline
Cyg\,X-1        &\ti{V1357\,Cyg}&  O9.7Iab  &  5.5996    & 0.25(1) &  33(9)   & \ti{(73.1,$+$3.1)}\\
\ti{1956$+$350} &\ti{HDE\,226868}&           &  74.7(10)  &         &  16(5)   &  2.5 \\ \hline
LMC\,X-3        &               &  B3Ve     &  1.70      & 2.3(3)  &          & \ti{(273.6,$-$32.1)}\\
\ti{0538$-$641} &               &           & 235(11)    &         & 5.6--7.8 &  55 \\ \hline
LMC\,X-1        &               & O7--9III  &  4.22      & 0.14(5) &          & \ti{(280.2,$-$31.5)}\\
\ti{0540$-$697} &               &           & 68(8)      &         &          &  55 \\ \hline
\hline
XN Mon 75         &\ti{V616\,Mon}&  K4 V   & 0.3230     & 2.83-2.99      & 0.53--1.22 & \ti{(210.0,$-$6.5)} \\
\ti{A\,0620$-$003}&\ti{N Mon 1917}& 0.057--0.077 & 443(4)     & 37--44$^{(\star)}$ &  9.4--15.9 & 0.66--1.45 \\ \hline
XN Oph 77       &\ti{V2107\,Oph}&  K3 V     & 0.5213     & 4.44--4.86 &  0.3--0.6 & \ti{(358.6,$+$9.1)}\\
\ti{H\,1705$-$250}&             &           & 420(30)    &   60--80   &  5.2--8.6 &  5.5: \\ \hline
XN Vul 88       &\ti{QZ\,Vul}    &  K5 V   & 0.3441     &  4.89--5.13 & 0.17--0.97  & \ti{(63.4,$-$3.1)}\\
\ti{GS\,2000$+$251}&            & 0.030--0.054 & 520(16)    &  43--74   &  5.8--18.0  &  2   \\ \hline
XN Cyg 89       &\ti{V404\,Cyg}&  K0 IV     & 6.4714     & 6.02--6.12  & 0.57--0.92  & \ti{(73.2,$-$2.2)}\\
\ti{GS\,2023$+$338}&\ti{N Cyg 1938, 1959}& 0.055--0.065 & 208.5(7)   &   52--60    & 10.3--14.2   & 2.2--3.7    \\ \hline
XN Mus 91       &               &  K5 V     & 0.4326     & 2.86--3.16 & 0.41--1.4 & \ti{(295.0,$-$6.1)}\\
\ti{GS\,1124$-$683}&           &  0.09--0.17 & 406(7)     &   54--65   & 4.6--8.2  & 3.0    \\ \hline
XN Per 92       &               &  M0 V   & 0.2127(7)  & 1.15--1.27 & 0.10-0.97 & \ti{(197.3,$-$11.9)}\\
\ti{GRO\,J0422$+$32}&           &  0.029--0.069 & 380.6(65)  & 28--45   & 3.4--14.0 &      \\ \hline
XN Sco 94       &               &  F5-G2    & 2.6127(8)  & 2.64--2.82 & 1.8--2.5 & \ti{(345.0,$+$2.2)}\\
\ti{GRO\,J1655$-$40}&           &  0.33--0.37 & 227(2)     &   67--71   & 5.5--6.8 & 3.2   \\ \hline
XN               &\ti{MX 1543-475}&  A2 V     & 1.123(8)   & 0.20--0.24 & 1.3--2.6  &  \ti{(330.9,$+$5.4)} \\
\ti{4U 1543$-$47}&                 &           & 124(4)     &  20-40    & 2.0--9.7  &  9.1(11)  \\ \hline
\hline
\end{tabular}
\end{center}
($\star$) A much higher inclination for A0620 has been claimed by Haswell et al. (1993) of up to i=70.
In this case, the lower limits on the component masses would be $M_X>3.8$ and $M_{opt}>0.22$.
\end{table}

\def\arraystretch{1.4}

\newpage


\begin{thebibliography}{DUM}
\bibitem[Bailyn et al. 1998]{ref:1}
   Bailyn, C. D., Jain, R. K., Coppi, P., \& Orosz, J. A. 1998, ApJ, 499, 367
\bibitem[Bailyn et al. 1998]{ref1}
   Barnes, C.A. 1995, Nucl. Phys. A, 588, 295c
\bibitem[aaaa]{ref:2}
   Bethe, H. A., \& Brown, G. E. 1998, ApJ, 506, 780
\bibitem[aaaa]{ref:3}
   Bethe, H. A., \& Brown, G. E. 1999, ApJ, accepted
\bibitem[aaaa]{ref:5-1}
   Brown, G. E. 1995, ApJ, 440, 270
\bibitem[aaaa]{ref:5}
   Brown, G. E. 1997, Phys. Bl., 53, 671
\bibitem[aaaa]{ref:6}
   Brown, G. E., \& Bethe, H. A. 1994, ApJ, 423, 659
\bibitem[aaaa]{ref:7-0}
   Brown, G. E., Lee, H. K., \& Wijers, R. A. M. J. 1999, in preparation
\bibitem[aaaa]{ref:7}
   Brown, G. E., Weingartner, J. C., \& Wijers, R. A. M. J. 1996, ApJ, 463, 297
\bibitem[aaaa]{ref:8}
    Callanan, P.J., et al. 1996, ApJ, 461, 351
\bibitem[aaaa]{ref:9}
    Cappellaro, E., et al. 1997, A\&A, 322, 431
\bibitem[aaaa]{ref:10}
    Chen, W., Shrader, C. R., \& Livio, M. 1997, ApJ, 491, 312
\bibitem[aaaa]{ref:10-1}
    Chevalier, R.A. 1993, ApJ, 411, L33
\bibitem[aaaa]{ref:10-2}
    Chevalier, R.A. 1996, ApJ, 459, 322
\bibitem[aaaa]{ref:11}
    Cordes, J.M., \& Chernoff, D.F. 1998, ApJ, 505, 315
\bibitem[aaaa]{ref:11-2}
    de Kool, M., van den Heuvel, E.P.J., \& Pylyser, E. 1987, A\&A, 183, 47
\bibitem[aaaa]{ref:12-1}
    Ergma, E., \& van den Heuvel, E.P.J. 1998, A\&A, 331, L29
\bibitem[aaaa]{ref:13}
    Haswell, C.A., Robinson, E.L., Horne, K., Stiening, R.F., 
   \& Abbott, T.M.C. 1993, ApJ, 411, 802
\bibitem[aaaa]{ref:14}
    Heap, S.R., \& Corcoran, M.F. 1992, ApJ, 387, 340
\bibitem[aaaa]{ref:14-2}
    Kaper, L., Lamers, H.J.G.L.M., van den Heuvel, E.P.J., 
    \& Zuiderwijk, E.J. 1995, A\&A, 300, 446
\bibitem[aaaa]{ref:14-3}
    Khaliullin, K.F., Khaliullina, A.I., \& Cherepashchuk, A.M.
    1984, Sov. Astron. Lett., 10, 250
\bibitem[aaaa]{ref:15-0}
    Langer, N. 1989, A\&A, 220, 135
\bibitem[aaaa]{ref:15-1}
    MacFadyen, A., \& Woosley, S. E. 1999, ApJ, accepted
\bibitem[aaaa]{ref:16-0}
    Marchenko, S.V., Moffat, A.F.J., \& Koenigsberger, G. 1994,
    ApJ, 422, 810  
\bibitem[aaaa]{ref:16-1}
    Mineshige, S., Nomoto, K., \& Shigeyama, T. 1993, A\&A, 267, 95
\bibitem[aaaa]{ref:16-2}
    Mineshige, S., Nomura, H., Hirose, M., Nomoto, K., \& Suzuki, T.
    1997, ApJ, 489, 227
\bibitem[aaaa]{ref:16-3}
    Moffat, A.F.J., \& Robert, C. 1994, ApJ, 421, 310
\bibitem[aaaa]{ref:16}
    Orosz, J. A., et al. 1998, ApJ, 499, 375
\bibitem[aaaa]{ref:20}
    Portegies Zwart, S. F., Verbunt, F., \& Ergma, E. 1997, A\&A, 321, 207
\bibitem[aaaa]{ref:24-0}
    Rubin, B.C., et al. 1996, ApJ, 459, 259
\bibitem[aaaa]{ref:24}
    Ruffert, M. 1994, ApJ, 427, 351
\bibitem[aaaa]{ref:25}
    Ruffert, M., \& Arnett, D. 1994, A\&AS, 106, 505
\bibitem[aaaa]{ref:26}
    St.-Louis, N., Moffat, A.F.J., Lapointe, L., Efimov, Y.S.,
    Shakhovskoy, N.M., Fox, G.K., \& Piirola, V. 1993, ApJ, 410, 342
\bibitem[aaaa]{ref:26-2}
    Schaller, G., Schaerer, D., Meynet, G., \& Maeder, A. 1992,
    A\&AS, 96, 269
\bibitem[aaaa]{ref:31}
   Timmes, F. X., Woosley, S. E., \& Weaver, T. A. 1996, ApJ, 457, 834
\bibitem[aaaa]{ref:31-1}
    van den Heuvel, E. P. J. 1995, J. Astrophys. Ast., 16, 255
\bibitem[aaaa]{ref:33}
    van den Heuvel, E. P. J., \& Habets, G.M.H.J. 1984, Nature, 309, 598
\bibitem[aa]{ref:34-2}
    Verbunt, F., \& van den Heuvel, E. P. J. 1995, in {\it X-ray Binaries},
    Cambridge Univ. Press, Eds. W. H. G. Lewin, J. van Paradijs \&
    E. P. J. van den Heuvel, p. 457
\bibitem[aa]{ref:34-3}
    Verbunt, F., \& Zwaan, C. 1981, A\&A 100, L7
\bibitem[aaaa]{ref:35}
    Weaver, T. A., \& Woosley, S. E. 1993, Phys. Rept., 227, 65
\bibitem[aaaa]{ref:36}
    Weaver, T. A., Zimmerman, G. B., \& Woosley, S. E. 1978, ApJ, 225, 1021
\bibitem[aaaa]{ref:36-1}
    Wellstein, S., \& Langer, N. 1999, to be published
\bibitem[aaaa]{ref:37}
    Wijers, R.A.M.J. 1996, Evolutionary Processes in Binary Stars,
  327-344, Kluwer Acad. Publ., Eds R.A.M.T. Wijers et al.
\bibitem[aaaa]{ref:38}
    Woosley, S. E. 1993, ApJ, 405, 273
\bibitem[aaaa]{ref:39}
    Woosley, S. E., Langer, N., \& Weaver, T. A. 1995, ApJ, 448, 315
\bibitem[aaaa]{ref:40}
    Woosley, S. E., \& Weaver, T. A. 1995, ApJS, 101, 181
\end{thebibliography}
\end{document}